\documentclass[a4paper]{VisionStyle}
\usepackage{psfig}

\def\gsimeq
{\hbox{\raise0.5ex\hbox{$>\lower1.06ex\hbox{$\kern-1.07em{\sim}$}$}}}
\def\lsimeq
{\hbox{\raise0.5ex\hbox{$<\lower1.06ex\hbox{$\kern-1.07em{\sim}$}$}}}

\begin{document}

\title{FIRST RESULTS FROM A {\it XMM-NEWTON} SURVEY OF A DISTANCE-LIMITED (D$<$22 Mpc) SAMPLE OF 
SEYFERT GALAXIES: I- THE AGNS}

\author{
M.\,Cappi\inst{1}, G.\,Di Cocco\inst{1}, F. Panessa\inst{1,7}, L. Foschini\inst{1}, M. Trifoglio\inst{1}, F. Gianotti\inst{1}, J. Stephen\inst{1}, L. Bassani\inst{1}, M. Dadina\inst{1}, A. Comastri\inst{2}, R. Della Ceca\inst{3}, A.V. Filippenko\inst{4}, L.C. Ho\inst{5}, K. Makishima\inst{6}, G. Malaguti\inst{1}, J. Mulchaey\inst{5}, G.G.C. Palumbo\inst{7,8}, E. Piconcelli\inst{1,7}, W. Sargent\inst{9}, K. Weaver\inst{10}, G. Zamorani\inst{2}
}

\institute{
IASF-CNR, Sezione di Bologna (formerly ITeSRE-CNR), Via Gobetti 101, 40129, Bologna, Italy
\and
Osservatorio Astronomico di Bologna, Via Ranzani 1, 40127, Bologna, Italy
\and
Osservatorio Astronomico di Brera, via Brera 28, 20121, Milano, Italy
\and
Department of Astronomy, University of California, Berkeley, CA 94720-3411, USA
\and
The Observatories of the Carnegie Institution of Washington, 813 Santa Barbara Street, Pasadena, CA 91101, USA 
\and
Department of Physics, University of Tokyo, 7-3-1 Hongo Bunkyoku, Tokyo 113-0033, Japan
\and
Universit\`a di Bologna, Dipartimento di Astronomia, via Ranzani 1, 40127 Bologna, Italy
\and
Agenzia Spaziale Italiana (ASI), Viale Liegi 26, 00198 Rome, Italy
\and
Palomar Observatory, MS 105-24, California Institute of Technology, Pasadena, CA 91125, USA
\and
NASA/Goddard Space Flight Center, Code 662, Greenbelt, Maryland 20771, USA
}

\maketitle 

\begin{abstract}

We report here preliminary results from a survey of nearby Seyfert galaxies 
using the EPIC CCDs on board {\it XMM-Newton}. 
The total sample consists of 28 Seyfert galaxies, and constitute a 
well-defined, complete ($B_{T}$$\lsimeq$12.5 mag), and volume-limited (D$<$22 Mpc) sample of 
Seyfert galaxies in the northern ($\delta > 0^\circ$) hemisphere.
The survey has been initiated in June, 2001, and we report here the results 
for the 6 objects analyzed so far, namely: NGC3185, NGC3486, NGC3941, NGC4138, NGC4565, 
and NGC5033. The main goal of this survey is to obtain a better and unbiased 
understanding of the ``typical'' Seyfert X-ray spectrum (e.g. the distribution of their 
absorption column density) in the local Universe. 
This is crucial to verify the predictions and, thus, to validate unified 
models, and is a fundamental parameter of synthesis models for the X-ray background. 
A companion poster (paper II: L. Foschini et al., these proceedings)  
illustrates how this survey will also allow a comprehensive spectral study of the
brightest (highest-luminosity) off-nuclear sources in the galaxies.

\keywords{Galaxies: active - X-rays: galaxies - Mission: XMM-Newton}
\end{abstract}

\section{Framework}

Important emphasis in research on Active Galactic Nuclei (AGNs) in recent 
years has been in the area of unified models (e.g. Antonucci 1993).
These models try to explain the observed 
differences between broad (Seyfert 1's-like) and narrow (Seyfert 2's-like) 
optical emission-line AGNs by invoking obscuration (from an optically and
geometrically thick torus) and viewing angle effects rather than intrinsic 
physical differences.
A privileged energy band for such studies is the hard (E$>$ 2 keV) X-ray band, 
where continuum photons are less affected by absorption and 
where imprinted features (e.g. the photoelectric absorption cut-off, the 
fluorescent FeK line) allow to probe the physics of the AGNs' nuclear regions, 
and their surroundings.

While there is increasing consensus on the validity of these models, 
at least for those that unify nearby ``classical'' Seyferts, questions remain 
on the ``typical'' Seyfert galaxy X-ray spectrum. This is particularly 
true for lower-luminosity objects and for the detailed relationship between 
their X-ray versus optical appearance and classification.
In particular, several open questions remain, such as whether 
the X-ray spectra of all nearby Seyferts are consistent with unified models
and whether the X-ray spectra of low-luminosity Seyferts are consistent 
with that of higher luminosity ones.
Also, precise knowledge of the distribution of column densities and 
the geometry of the putative molecular torus, of the origin of FeK emission 
lines, and of the shape of the X-ray luminosity 
function of local Seyfert galaxies once extended to its fainter end, 
is still lacking.

Answers to the above questions are also crucial to put tighter 
constraints on synthesis models of the XRB and to understand the evolution 
of ``Seyfert-like'' AGNs with cosmic time (e.g. Comastri et al. 1995, 
Gilli et al. 2000). These models 
usually assume the validity of unified models and suggest that 
a combination of unabsorbed (Seyfert 1-type) and absorbed (Seyfert 2-type) 
AGNs may account for most of the X-ray background (XRB). 
But in fact, major uncertainties remain in the assumed $N_{\rm H}$ 
distributions, luminosity functions of both types of objects and their 
evolution with cosmic time.

The definition of the spectral properties of a complete and, {\it bona fide}, 
unbiased sample of nearby Seyfert galaxies is, thus, desired 
to address both the issues of unified models of AGNs and synthesis models of the XRB.
 

\section{Previous and Current Studies}

Hard X-ray samples of nearby Seyfert galaxies available to date (e.g. from {\it GINGA}, {\it ASCA} and 
{\it BeppoSAX}; Smith \& Done 1996, Turner et al. 1998, Bassani et al. 199) have been essentially 
biased towards the most X-ray luminous, 
and less absorbed AGNs\footnote{But see the pioneering works by Maiolino et al. (1998) 
and Risaliti, Maiolino \& Salvati (1999), and references therein.}.
Moreover, the poor spatial resolution of the pre-{\it Chandra} and pre-{\it XMM-Newton} 
telescopes left unresolved issues such as a proper estimate of 
stellar/starburst processes and/or X-ray binaries contributions to the total 
X-ray emission of low-luminosity (L$_{\rm 2-10 keV}$ $\lsimeq$ 10$^{40}$erg/s) 
Seyfert galaxies.
Significant improvements have been obtained by Ho et al. (2001) 
using the ACIS CCDs on board {\it Chandra}, who can address for the first time 
the X-ray properties of LLAGNs (with 2-10 keV luminosities ranging from less than 10$^{38}$ to 
10$^{41}$ erg/s) in nearby Seyferts, LINERs and LINERs/HII transition nuclei.
These authors detected a compact, pointlike nuclear source in $\sim$62\% of their 
initial sample of 24 (out of 41) nearby galaxies.
However, the brief ``snapshots'' exposures 
used in the Chandra survey will make it difficult to obtain detailed spectral information 
at energies greater than 2 keV.

Given the above arguments, we have initiated an {\it XMM-Newton} X-ray survey
of all known northern Seyfert galaxies with D$<$22 Mpc using the 
Revised-Ames Catalog of Bright Galaxies (Sandage \& Tamman 1981) spectroscopically 
classified by Ho, Filippenko \& Sargent (1997) according the classical definition 
of Seyfert galaxies (Veilleux \& Osterbrock, 1987). 
The final sample contains 28 galaxies (see Table 1) and is the deepest and most 
complete local sample of Seyfert galaxies.
The goal of this survey is to obtain a better, and unbiased, understanding of 
their ``typical'' X-ray spectrum.
This high-throughput (especially for energies $>$2 keV) {\it XMM-Newton} survey is 
complementary to the arcsecond-resolution {\it Chandra} survey performed by L. Ho, 
E.D. Feigelson, and collaborators.

\section{The {\it XMM-Newton} Sample}

Our team has been awarded $\sim$ 250 ks of EPIC guarantee time, and exposures 
of 5, 10, 15 and 20 ks (with a typical value of 10 ks) were requested in order to obtain 
at least $\sim$ 1000 source counts\footnote{Expected count-rates were obtained from the 
observed 2-10 keV fluxes from Polletta et al. (1996) and Serlemitsos et al. (1997) and, 
when not available, from the L$_{2-10 \rm {keV}}$-L$_{\rm H_{\alpha}}$ correlation of 
Serlemitsos et al. (1997).}.

To date, observations have been performed for about half of the galaxies, and 
data are available for 6 of them. Here we report on the preliminary 
results obtained for this subsample.
The data were processed and screened with the Science Analysis Software 
(SAS; version 5.2) using standard procedures. The latest known calibration files and 
response matrices released by the EPIC team have been used.
Data from MOS1/2 and PN were combined in the spectral analysis.

\begin{table}[bht]
\caption{Distance-Limited (D$<$22 Mpc) Sample of Seyfert Galaxies}
\begin{center}
\leavevmode
\footnotesize
\begin{tabular}{cccc}
\hline
\hline
\multicolumn{1}{c}{NGC} &
\multicolumn{1}{c}{Distance} &
\multicolumn{1}{c}{B} &
\multicolumn{1}{c}{Seyfert} \\
\multicolumn{1}{c}{Galaxy} &
\multicolumn{1}{c}{(Mpc)} &
\multicolumn{1}{c}{(mag)} &
\multicolumn{1}{c}{Type} \\
\hline
  3031 &   1.4 &  7.89  & S1.5\\ 
  4395 &   3.6 & 10.64  & S1.8\\   
  4258 &   6.8 &  9.10  & S1.9\\   
  3486 &   7.4 & 11.05  & S2  \\   
  5194 &   7.7 &  8.96  & S2  \\   
  1058 &   9.1 & 11.82  & S2  \\   
  4565 &   9.7 & 10.42  & S1.9\\   
  4725 &  12.4 & 10.11  & S2  \\  
  2685 &  16.2 & 12.12  & S2  \\  
  4168 &  16.8 & 12.11  & S1.9\\  
  4388 &  16.8 & 11.76  & S1.9\\  
  4472 &  16.8 & 9.37   & S2  \\  
  4477 &  16.8 & 11.38  & S2  \\  
  4501 &  16.8 & 10.36  & S1.9\\ 
  4579 &  16.8 & 10.48  & S1.9\\  
  4639 &  16.8 & 12.24  & S1.0\\  
  4698 &  16.8 & 11.46  & S2  \\  
  3982 &  17.0 & 11.78  & S1.9\\  
  4051 &  17.0 & 10.83  & S1.2\\  
  4138 &  17.0 & 12.16  & S1.9\\  
  5033 &  18.7 & 10.75  & S1.5\\  
  3941 &  18.9 & 11.25  & S2  \\  
   676 &  19.5 & 10.50  & S2  \\  
  4151 &  20.3 & 11.50  & S1.5\\  
  3079 &  20.4 & 11.54  & S2  \\  
  3227 &  20.6 & 11.10  & S1.5\\  
  3185 &  21.3 & 12.99  & S2  \\ 
  5273 &  21.3 & 12.44  & S1.5\\  
\hline
\hline
\end{tabular}
\end{center}
\end{table}
\normalsize

\section{Preliminary Results}

Two of the earliest observations (NGC5033 and NGC3941) suffer 
from contamination by soft-proton flares which increase significantly the background
(see Fig. 1, left panel) and, thereby, complicate/limit the scientific results.
Nevertheless, examination of the images in Fig. 1 and 2 clearly illustrate the 
capabilites of this survey. Despite the short exposures, we are able to detect and 
distinguish the nuclei from the off-nuclear sources in the galaxies.
The high throughput also allows us to perform 
a detailed spectral analysis of most sources. Three examples of such spectra 
are shown in Fig. 3. 
In these spectra, MOS1/2 and PN data were fitted simultaneously and with simple 
models (a single absorbed power-law for NGC5033 and NGC4565, and an absorbed plus 
scattered power-law for NGC4138) to illustrate the statistics available, as well 
as to highlight the FeK features in the residuals for NGC5033 and NGC4138.

\begin{figure*}[!ht]
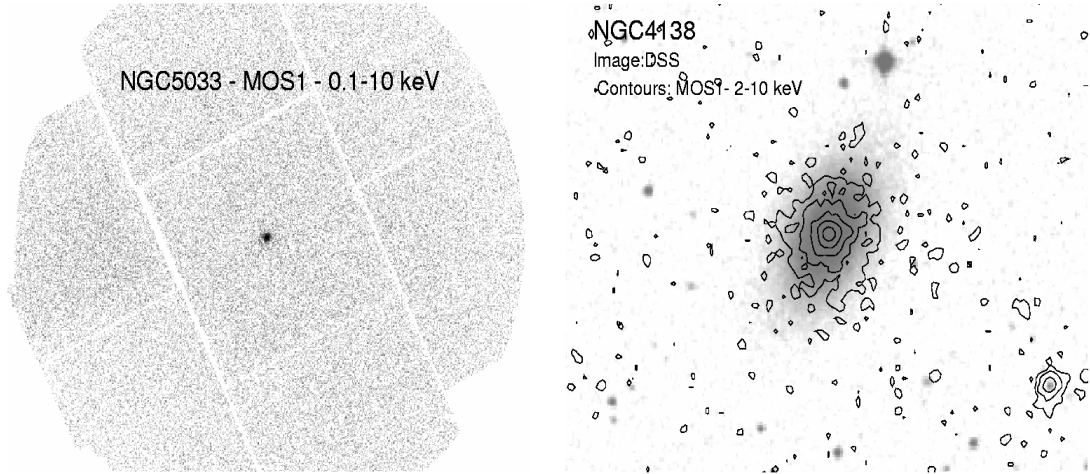

\begin{center}
\parbox{8cm}{
\psfig{file=mcappi-C2_fig1.ps2,width=7cm,height=6.3cm,angle=0}}
\hspace{0.1cm} \
\parbox{8cm}{
\psfig{file=mcappi-C2_fig2.ps2,width=7cm,height=6.3cm,angle=0}}
\end{center}
\caption{(Left panel) MOS1 0.1-10 keV image of NGC5033. Despite the fact that this 
observation was contaminated by soft-proton flares (as demonstrated by the high background level), 
the source is clearly detected. (Right panel) MOS1 emission contours between 2-10 keV 
overlaid on the DSS image for NGC4138. These images illustrate well the cases where the 
sources are relatively bright, i.e. with F$_{\rm 2-10 keV}$$\gsimeq$10$^{-12}$erg cm$^{-2}$ s$^{-1}$.}
\end{figure*}

\begin{figure*}[ht]
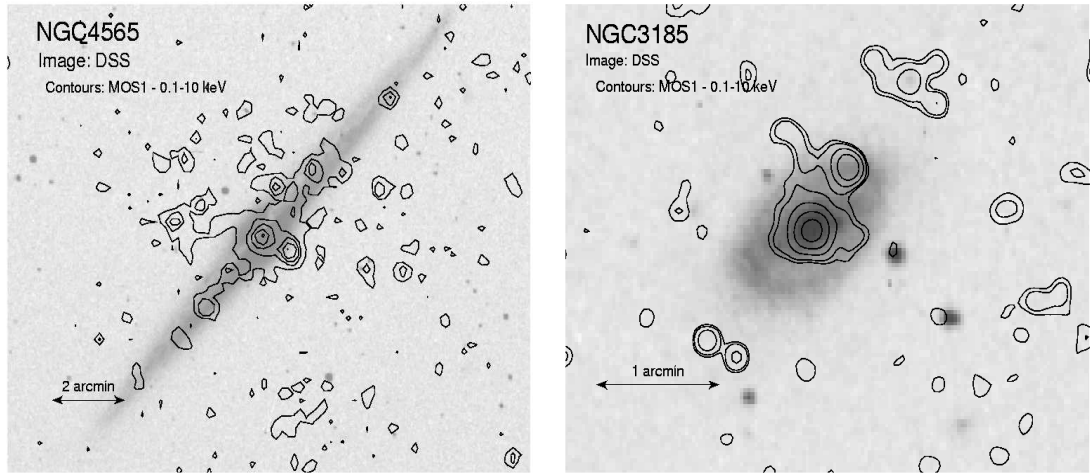

\begin{center}
\parbox{8cm}{
\psfig{file=mcappi-C2_fig3.ps2,width=7cm,height=6.3cm,angle=0}}
\hspace{0.1cm} \
\parbox{8cm}{
\psfig{file=mcappi-C2_fig4.ps2,width=7cm,height=6.3cm,angle=0}}
\end{center}
\caption{MOS1 (0.1-10 keV) contours overlaid on the DSS plates images of NGC4565 (Left) and 
NGC3185 (Right).}
\end{figure*}

Preliminary spectral fits of the nuclear sources 
are summarized in Table 2. A discussion of the results in terms of ``sample 
properties'' is clearly premature here and will be presented in a forthcoming 
paper once other observations will be available. 
Nevertheless, these early results highlight the importance of this {\it XMM-Newton} 
survey. They confirm in fact that we shall be able to measure for the first
time with good precision the absorption column 
densities, the spectral shapes, the soft (0.5-2 keV) and hard (2-10 keV) 
fluxes/luminosities, and also the FeK line parameters (at least for the brightest targets) for 
a significant number of Seyferts with luminosities down to $\sim$10$^{39}$erg/s.
We can also stress here some source-by-source interesting results that came out 
from this early analysis:
i) the detection of an FeK line in NGC5033 and NGC4138 with EW$\sim$100-200 eV; 
ii) the detection of an absorbed component in NGC4138 and, possibly, in NGC3486; 
iii) the evidence of extended soft X-ray emission in NGC4138 and NGC4565, and iv) the 
presence of luminous off-nuclear sources in 4 out of 6 galaxies\footnote{See the companion 
paper by Foschini et al. (these proceedings) which illustrates well 
the capabilities of this survey to the study of ultraluminous X-ray sources in 
nearby galaxies.}.

Last but not least, data from this survey will also be used to extend the spectroscopic 
survey of serendipitous sources initiated by Piconcelli et al. (these proceedings). 
In total, we expect that $\gsimeq$ 20 serendipitous sources should be detected 
with EPIC in hard (2-10 keV) X-rays and give enough photons for a proper spectral 
analysis (see Piconcelli et al. for more details).


\begin{table*}[!ht]
\begin{center}
\caption{Nuclear hard X-ray properties}
\leavevmode
\footnotesize
\begin{tabular}{cccccccc}
\hline
\hline
\multicolumn{1}{c}{Name} &
\multicolumn{1}{c}{Sey} &
\multicolumn{1}{c}{Exp.} &
\multicolumn{1}{c}{$N_{\rm H}$} &
\multicolumn{1}{c}{$\Gamma_{\rm 2-10 keV}$} &
\multicolumn{1}{c}{F$_{\rm 2-10 keV}$} &
\multicolumn{1}{c}{L$_{\rm 2-10 keV}$} &
\multicolumn{1}{c}{X-ray Class$^*$} \\
\multicolumn{1}{c}{} &
\multicolumn{1}{c}{Type} &
\multicolumn{1}{c}{(ks)} &
\multicolumn{1}{c}{cm$^{-2}$} &
\multicolumn{1}{c}{} &
\multicolumn{1}{c}{$\times$10$^{-14}$cgs} &
\multicolumn{1}{c}{$\times$10$^{39}$ erg/s} &
\multicolumn{1}{c}{} \\
\hline
  NGC3185 &  S2  &	13	&$\equiv$$N_{\rm Gal}$&	2.1$\pm$0.4&	1.4&	0.9&	I\\ 
  NGC3486 &  S2  &	5.5	&$>$10$^{23}$&	2.2$\pm$0.3&	5&	$>$1&	II-III\\   
  NGC3941 &  S2  &	7	&$\equiv$$N_{\rm Gal}$&	2.3$\pm$0.4&	2&	0.7&	II-III\\  
  NGC4138 &  S1.9&	14	&8.2$\pm$0.7$\times$10$^{22}$&	1.6$\pm$0.1&	500&	400&	I\\  
  NGC4565 &  S1.9&	14	&$\equiv$$N_{\rm Gal}$&	1.7$\pm$0.2&	20&	15&	II\\   
  NGC5033 &  S1.5&	7.5	&$\equiv$$N_{\rm Gal}$&	1.7$\pm$0.2&	280&	150&	I\\  
\hline
\hline
\end{tabular}
\end{center}
$^*$X-ray morphology class following the criteria given in Ho et al. (2001). This classification 
indicates 4 classes of X-ray sources where (I) the nucleus dominates the total X-ray emission; 
(II) the nucleus is comparable in brightness to off-nuclear sources in the galaxy; (III) the nucleus is 
embedded in diffuse emission; and (IV) the nucleus is absent.
\end{table*}

\par\noindent

\begin{figure}[!]
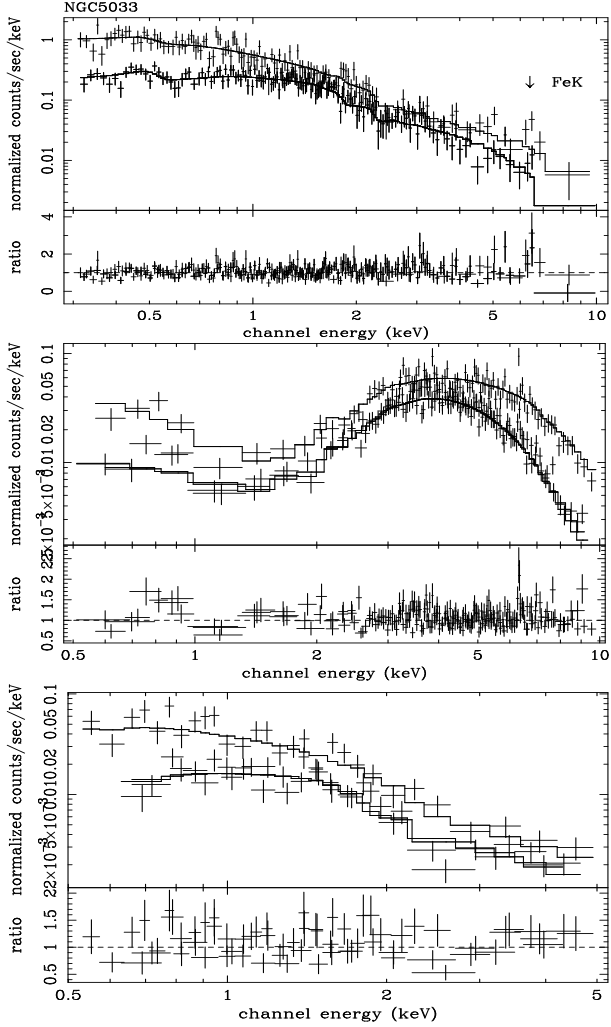

\begin{center}
\psfig{file=mcappi-C2_fig5.ps,width=8cm,height=4.5cm,angle=-90}
\psfig{file=mcappi-C2_fig6.ps,width=8cm,height=4.5cm,angle=-90}
\psfig{file=mcappi-C2_fig7.ps,width=8cm,height=4.5cm,angle=-90}
\end{center}
\caption{MOS and PN spectra of NGC5033 (top), NGC4138 (middle) and NGC4565 (bottom). 
The models used are a single absorbed power-law model for NGC5033 and NGC4565, and 
an absorbed plus scattered component for NGC4138.    
Residuals are plotted in the form of data to model ratios. FeK lines are clearly 
detected in NGC5033 and NGC4138. These spectra illustrate well the kind of spectra 
we expect from this survey.}
\end{figure}

\par\noindent

\vspace{3truecm}
\acknowledgements

We are grateful to Matteo Guainazzi and Silvano Molendi 
for helpful discussions on data reduction procedures.
We gratefully acknowledge partial support funding from the Italian 
Space Agency (ASI).   
This work is based on observations obtained with {\it XMM-Newton}, an ESA 
science mission with instruments and contributions directly 
funded by ESA Member States and the USA (NASA). 
This research has made use of the NASA/IPAC Extragalactic Database (NED) which is operated 
by the Jet Propulsion Laboratory, California Institute of Technology, under contract with the 
National Aeronautics and Space Administration.

\end{document}